\documentclass[useAMS,usenatbib,usegraphicx]{mn2e}

\usepackage{amsmath,amssymb,color,latexsym}

\begin{document}

%%%%%%%%%%%%%%%%%%%%%%%%%%%%%%%%%%%%%%%%%%%%%%%%%%%%%%%%%%%%%%%%%%%%%%%%%%
%%%%%%%%%%%%%%%%%%%%%%%%%%%%%%%%%%%%%%%%%%%%%%%%%%%%%%%%%%%%%%%%%%%%%%%%%%
%%%%%%%%%%%%%%%%%%%%%%%%%%%%%%%%%%%%%%%%%%%%%%%%%%%%%%%%%%%%%%%%%%%%%%%%%%
\title[Gravitational Waves from Neutrino-Driven GRBs]{Gravitational
Wave Background from
Neutrino-Driven Gamma-Ray Bursts}
%%%%%%%%%%%%%%%%%%%%%%%%%%%%%%%%%%%%%%%%%%%%%%%%%%%%%%%%%%%%%%%%%%%%%%%%%%
%%%%%%%%%%%%%%%%%%%%%%%%%%%%%%%%%%%%%%%%%%%%%%%%%%%%%%%%%%%%%%%%%%%%%%%%%%
%%%%%%%%%%%%%%%%%%%%%%%%%%%%%%%%%%%%%%%%%%%%%%%%%%%%%%%%%%%%%%%%%%%%%%%%%%

\author[Hiramatsu et al.]{
Takashi Hiramatsu$^{1}$\thanks{E-mail:hiramatsu-at-utap.phys.s.u-tokyo.ac.jp}, 
Kei Kotake$^{2}$, 
Hideaki Kudoh$^{1}$ and
Atsushi Taruya$^{1}$\\
$^1$Department of Physics, School of Science, University of Tokyo,
7-3-1 Hongo, Bunkyo, Tokyo 113-0033, Japan\\
$^2$Science \& Engineering, Waseda University, 3-4-1 Okubo, Shinjuku,
Tokyo, 169-8555, Japan}

\date{Accepted 0000 December 00. Received 0000 December 00; in original form 0000 October 00}

\pagerange{\pageref{firstpage}--\pageref{lastpage}} \pubyear{0000}

\maketitle

\label{firstpage}

%%%%%%%%%%%%%%%%%%%%%%%%%%%%%%%%%%%%%%%%%%%%%%%%%%%%%%%%%%%%%%%%%%%%%%%%%%
\begin{abstract}
We discuss the gravitational wave background (GWB) 
from a cosmological 
population of gamma-ray bursts (GRBs).
Among various emission 
mechanisms for the gravitational waves (GWs), we pay a particular 
attention to the 
vast anisotropic neutrino emissions from the accretion disk around the
black hole formed after the so-called failed supernova explosions. 
The produced GWs by such mechanism are known as {\it burst with memory},  
which could dominate over the low-frequency regime below $\sim10$Hz. 
To estimate their amplitudes, we derive general analytic formulae for 
gravitational waveform from the axisymmetric jets.
Based on the formulae, we first quantify the 
the spectrum of GWs from a single 
GRB. Then, summing up its cosmological population,
we find that the resultant value of the density parameter becomes 
roughly $\Omega_\mathrm{GW} \approx 10^{-20}$ 
over the wide-band of the low-frequency region, 
$f\sim 10^{-4}-10^1$Hz. The amplitude of GWB is sufficiently 
smaller than the primordial GWBs originated from an 
inflationary epoch and far below the detection limit. 
\end{abstract}
%%%%%%%%%%%%%%%%%%%%%%%%%%%%%%%%%%%%%%%%%%%%%%%%%%%%%%%%%%%%%%%%%%%%%%%%%%

%%%%%%%%%%%%%%%%%%%%%%%%%%%%%%%%%%%%%%%%%%%%%%%%%%%%%%%%%%%%%%%%%%%%%%%%%%
%%%%%%%%%%%%%%%%%%%%%%%%%%%%%%%%%%%%%%%%%%%%%%%%%%%%%%%%%%%%%%%%%%%%%%%%%%
\section{Introduction} 
\label{sec:intro}
%%%%%%%%%%%%%%%%%%%%%%%%%%%%%%%%%%%%%%%%%%%%%%%%%%%%%%%%%%%%%%%%%%%%%%%%%%
%%%%%%%%%%%%%%%%%%%%%%%%%%%%%%%%%%%%%%%%%%%%%%%%%%%%%%%%%%%%%%%%%%%%%%%%%%

The observation of gravitational waves (GWs) is one of the most important
missions to explore the trackless parts of the Universe. 
Several ground-based laser interferometers (TAMA300, LIGO, and GEO600)
are now operating and taking continuously the data at a frequency range, 
10Hz -- 10kHz, where 
rapidly-moving stellar objects accompanied with strong gravity, 
such as formations of neutron stars or black holes, are 
the most promising sources for GWs (See \citet{new} for a review). 
The Laser Interferometer Space 
Antenna, {\it LISA}\footnote{See http://lisa.jpl.nasa.gov/.} covering 
$10^{-4}-10^{-2}$ Hz will be launched near future, and 
moreover, the future space missions such as 
BBO\footnote{See http://universe.nasa.gov/program/bbo.html .}
or DECIGO \citep{DECIGO} target the frequency window around 0.1Hz. For
such interferometers, the primordial GWs generated during an inflationary 
epoch could be detected. Also, these interferometers 
are expected to be useful in determining the cosmic equation of the state 
\citep{TakahashiR}. Thus, low-frequency GWs detected via 
space interferometers may provide 
a powerful cosmological tool to probe the extremely early Universe 
\citep{Maggiore}. 
One important remark 
is that several astrophysical foregrounds around 
deci-Hertz band proposed so far are believed to be weak or resolvable 
\citep[e.g.,][]{Ferrari,Schneider,Farmer}.

Recently, however, \citet{Buo} pointed out the possibility that a
cosmological population of core-collapse supernovae could contaminate the 
inflationary GWs at the low-frequency range mentioned above, as a result
of the GW associated with the neutrino emissions. Their 
analysis suggests that even in a deci-hertz band, one may not simply neglect 
the contributions from the cosmological stellar objects, 
especially with the asymmetric emissions of neutrinos.  
The important remark is that GWs from neutrino-driven jets 
are generated by different emission mechanisms from those of 
the periodic GWs. 
For an energy flow originating from a burst, 
the GW amplitude suddenly rises from zero and settles down into 
a non-vanishing value after the burst. 
This phenomenon is called {\it burst with memory} \citep{BT}. 
It has been argued that a burst of neutrinos released 
anisotropically can generate a burst of GW accompanying with the memory 
effect. A cosmological contribution of such GW may therefore become 
significant at the low-frequency band rather than the 
high-frequency regime.

Among several candidates 
other than the core-collapse supernovae, 
GRBs are expected to have powerful asymmetric jets with the enormous 
emission of high energy neutrinos. It has been already pointed out
that relativistic jets of matter associated with GRBs can generate strong
GWs enough to be detected by the present laser interferometers 
\citep{Sago}. In fact, a data analysis searching for GWs associated with the 
very bright GRB030329 have been performed by LIGO group \citep{LIGO}, 
although they reported a null result.

Another important aspect is that these jets are thought to be
driven by thermal energy deposition due to the neutrino and
anti-neutrino annihilation into electrons and positrons, which occurs
primarily at the polar region of the black hole-accretion disk 
system, namely, the neutrino-driven GRBs 
\citep{pac,remes,popham,ruffajan,MacFadyen,asano} 
\citep[see,][for the MHD processes]
{proga,mizuno,taki}. Naively inferred from the energetics of 
core-collapse supernovae \citep{burohey,fry04,muller03,kotakesub}, 
anisotropic energy flows from the GRBs also become 
important sources of GWs, in addition to the anisotropic jets of the
matter itself \citep{Sago}. 
While a cosmological contribution of those GWs may not play a significant role 
due to the relatively rare event rate of the GRBs, 
a quantitative estimate of the shape and the amplitude of 
the spectrum still remains important issue to clarify 
the nature of gravitational wave background arising from the memory effect.  
We then wish to understand how the GW memory is generated 
and how much amount of the GWs is released 
from a cosmological population of GRBs in an analytic manner.

The purpose of this paper is to give an analytic estimate of 
the gravitational wave spectrum from the GW memory effect, particularly 
focusing on the neutrino-driven GRBs. 
In Sec.\ref{sec:gw-formalism}, after 
briefly describing the axisymmetric jet model,  
we derive a general analytic formula for gravitational waveform with
memory effect from the asymmetric relativistic jets, which is 
both applicable to the ultra-relativistic and the relativistic cases. 
Using this formula, in Sec.\ref{sec:gw-single}, 
we compute the spectrum of GWs from a 
single neutrino-driven GRB. Then, 
in Sec.\ref{sec:gw-background}, summing up a cosmological population of GRBs,  
we estimate the amplitude of 
gravitational wave background (GWB). Uncertainties of the 
model parameters are also taken into account when evaluating the GWB, 
and we find that 
the resultant GWB amplitude becomes rather small and would be difficult 
to detect by the present and future missions of GW observatories. 
Finally, summary and discussion are devoted to section \ref{sec:summary}.

%%%%%%%%%%%%%%%%%%%%%%%%%%%%%%%%%%%%%%%%%%%%%%%%%%%%%%%%%%%%%%%%%%%%%%%%%%
%%%%%%%%%%%%%%%%%%%%%%%%%%%%%%%%%%%%%%%%%%%%%%%%%%%%%%%%%%%%%%%%%%%%%%%%%%
\section{Gravitational Waves from relativistic Jets} 
\label{sec:gw-formalism}
%%%%%%%%%%%%%%%%%%%%%%%%%%%%%%%%%%%%%%%%%%%%%%%%%%%%%%%%%%%%%%%%%%%%%%%%%%
%%%%%%%%%%%%%%%%%%%%%%%%%%%%%%%%%%%%%%%%%%%%%%%%%%%%%%%%%%%%%%%%%%%%%%%%%%

In this paper, we specifically treat the GWs from 
the ultra-relativistic jets, particularly focusing on 
the neutrino-driven GRBs as a plausible GW source.
In this case, 
energy sources powering the GRBs rely on the neutrino emission from the 
accreting disk \citep[][for a review and references therein]{Piran}.
Recently, \citet{Seti}
performed the three-dimensional hydrodynamic simulations and obtained
the time evolution of the luminosity of neutrinos emitted from accretion
disks around hyper-accreting stellar-mass black hole. 
Using their results, \citet{Aloy} performed the two-dimensional
simulations to study the evolution of relativistic jets driven by
thermal energy deposition due to $\nu\overline{\nu}$-annihilation.
In next section, we adopt their fitting result for the neutrino
luminosity evolution to calculate the GW amplitude.  
Before estimating this, we discuss the GW memory and 
derive a general analytic formula for gravitational waveform from 
the axisymmetric relativistic jets.

According to the memory effect of GWs, 
the GW amplitude $h(t)$ jumps from zero 
to a non-vanishing value and it keeps maintaining the non-vanishing 
value even after the energy source of GWs disappears. 
\citet{Epstein} has derived a general formula of memory effect 
produced by the radial emission of neutrinos. 
The detectability of such effect was discussed by 
\citet{BT} through the ground-based laser interferometers.  
Recently, \citet{Segalis} studied the memory effect 
generated by a point particle whose velocity
changes via gravitational interactions with other objects. 
Further, \citet{Sago} estimated  the amplitude of GW memory 
from the GRB jets based on the internal shock model.   
Here, we consider the GWs from the neutrino-driven GRB and 
derive a useful analytic formula for
the axisymmetric emission of neutrinos.

\begin{figure}
 \centering
 \includegraphics[width=5cm]{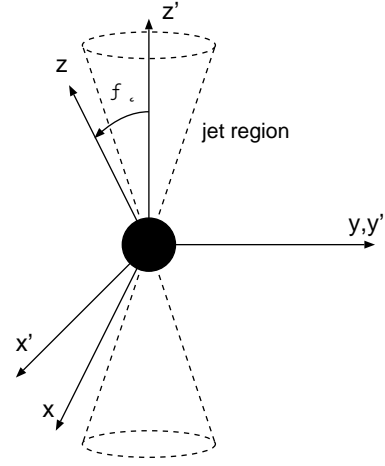}
 \caption{Source coordinate system ($x',y',z'$) and observer coordinate
 system ($x,y,z$). The observer resides at the distant point on the
 $z$-axis. The $z'$-axis coincides with the symmetric axis of the GRB 
 source. Inside the jet represented 
   by dashed lines, the luminosity of energy flow is assumed to 
   be uniformly distributed with the opening angle $\theta_{\rm open}$.}
 \label{fig:coordinate}
\end{figure}

Let us introduce the two coordinate systems shown in 
figure \ref{fig:coordinate};  
 the source coordinate system $(x',y',z')$ and 
the observer coordinate system $(x,y,z)$. 
In these coordinate systems, 
the $z'$-axis is chosen as jet axis, while the 
$z$-axis is set to a line-of-sight direction. 
Further, the origins of these two coordinate systems 
are set to the centre of a GRB object. 
Then, the viewing-angle denoted by $\xi$ is given by 
the angle between $z$- and $z'$-axis. 
For convenience, we assume that the $z$-axis 
lies on the $(x',z')$-plane. In this case, 
the two polarization states of GWs satisfying the 
transverse-traceless conditions become $h_+\equiv h_{xx}=-h_{yy}$ and 
$h_\times\equiv h_{xy}=h_{yx}$ in the observer coordinates and  
we obtain $h_\times=0$ (see below). 
Note that the sum of the squared amplitudes, 
$|h_+|^2+|h_\times|^2$, is invariant under the rotation about the 
$z$-axis. 
The geometrical setup shown in figure \ref{fig:coordinate} 
yields the following relation between the two polar
coordinate systems $(\theta,\phi)$ and $(\theta',\phi')$:
%%%%%%%%%%%%%%%%%%%%%%%%%%%%%%%%%%%%%%%%%%%%%%%%%%%%%%%%%%%%%%%%%%%%%%%%%%%
\begin{eqnarray}
&&   \sin\theta'\cos\phi'= \sin\theta\cos\phi\cos\xi +
    \cos\theta\sin\xi, 
\nonumber \\
&&   \sin\theta'\sin\phi'= \sin\theta\sin\phi, 
\nonumber \\
&&   \cos\theta'         = -\sin\theta\cos\phi\sin\xi + \cos\theta\cos\xi.
 \label{eq:trans}
\end{eqnarray} 
%%%%%%%%%%%%%%%%%%%%%%%%%%%%%%%%%%%%%%%%%%%%%%%%%%%%%%%%%%%%%%%%%%%%%%%%%%%

Since the GRBs are cosmological, one can safely neglect the time lag between the two points near the source region, i.e., $|t-t'|\ll r/c$
\citep{MJ}. Thus the amplitude of GWs observed by the distant observer may be written as 
%%%%%%%%%%%%%%%%%%%%%%%%%%%%%%%%%%%%%%%%%%%%%%%%%%%%%%%%%%%%%%%%%%%%%%%%%%%
\begin{eqnarray}
  && h_+(t) = \frac{2G}{c^4r}\int^{t-r/c}_{-\infty}\!dt'
          d\Omega'\,\psi(\Omega')\frac{dL}{d\Omega'}(\Omega',t'), \quad
  \cr
  && \psi(\Omega')  =  \frac{\beta^2\sin^2\theta}{1-\beta\cos\theta}\cos 2\phi,
\label{eq:angular}
\end{eqnarray}
%%%%%%%%%%%%%%%%%%%%%%%%%%%%%%%%%%%%%%%%%%%%%%%%%%%%%%%%%%%%%%%%%%%%%%%%%%%
where $\beta$ is the velocity of matter in a jet normalized by $c$, and
$\theta$ and $\phi$ are related to $\theta'$ and $\phi'$ through 
the relation (\ref{eq:trans}). The function $\psi(\Omega')$ represents 
the dependence of the amplitude on the direction of the energy flow
\citep{Segalis}. Note that the counter part of the amplitude, 
$h_\times$, is obtained 
just replacing $\cos 2\phi$ with $\sin 2\phi$ in 
equation (\ref{eq:angular}), which immediately yields 
$h_\times=0$ by integrating over the angle $\phi'$. 
The quantity $dL/d\Omega'$ is the neutrino luminosity per unit solid 
angle emitted from the source. For an axisymmetric neutrino emission, 
we consider the two axisymmetric jets whose luminosity is uniformly 
distributed with respect to the opening angle of jets. 
Denoting the opening angle by $\theta_{\rm open}$, we have   
%%%%%%%%%%%%%%%%%%%%%%%%%%%%%%%%%%%%%%%%%%%%%%%%%%%%%%%%%%%%%%%%%%%%%%%%%%%
\begin{eqnarray}
&& \frac{dL}{d\Omega} (\Omega,t)
    = \frac{E }{\Delta\Omega_{\rm jet}} \Gamma(t) 
\nonumber\\
&& \qquad 
   \times 
   \left[ \Theta\left(\frac{\theta_{\rm open}}{2}-\theta'\right)
   +\Theta\left(\theta'-\pi+\frac{\theta_{\rm open}}{2}\right)
   \right],
\label{eq:luminosity}
\end{eqnarray}
%%%%%%%%%%%%%%%%%%%%%%%%%%%%%%%%%%%%%%%%%%%%%%%%%%%%%%%%%%%%%%%%%%%%%%%%%%%
where the function $\Theta(\theta)$ is Heaviside step function.  
The quantity $E$ is the total isotropic emission energy, which is 
typically $E\simeq \gamma mc^2 \approx 10^{51}$ erg in the case of 
jets from relativistic matter, and
$E_{\nu} \la 10^{52}$ erg in the case of neutrino emission
\citep{popham}.
In equation (\ref{eq:luminosity}), 
$\Delta\Omega_{\rm jet}$ denotes the solid angle of two jets 
given by 
$\Delta\Omega_{\rm jet}\equiv 4\pi\{1-\cos(\theta_{\rm open}/2)\}$. 
The function $\Gamma(t)$ represents the time dependence of the
luminosity of the ejected matter or neutrinos, which will be given
explicitly.

Substituting equation (\ref{eq:luminosity}) into equation
(\ref{eq:angular}) and   
performing the Fourier transformation $2\pi if\widetilde{h}(f) =
\int_{-\infty}^{\infty}\!\dot{h}(t)e^{-2\pi ift}\,dt$, the
characteristic amplitude defined by $h_c(f)\equiv f\widetilde{h}(f)$ 
is given by 
%%%%%%%%%%%%%%%%%%%%%%%%%%%%%%%%%%%%%%%%%%%%%%%%%%%%%%%%%%%%%%%%%%%%%%%%%%%
\begin{eqnarray}
 h_c(f) &=& \frac{G}{\pi c^4r}\frac{\Psi}{\Delta\Omega_{\rm jet}}
          E|\widetilde{\Gamma}(f)| \label{eq:h_c}\\
        &=& 8.5\times 10^{-21}\left(\frac{10{\rm kpc}}{r}\right)
          \left(\frac{E}{10^{52}{\rm erg}}\right)
          \frac{\Psi}{\Delta\Omega_{\rm jet}}|\widetilde{\Gamma}(f)|.\nonumber 
\end{eqnarray}
%%%%%%%%%%%%%%%%%%%%%%%%%%%%%%%%%%%%%%%%%%%%%%%%%%%%%%%%%%%%%%%%%%%%%%%%%%%
Here, the function $\tilde{\Gamma}(f)$ is the Fourier transform of the 
function $\Gamma$. The quantity $\Psi$ is the angular integral 
of the function $\psi(\Omega')$, which depends on 
the parameters, $\theta_{\rm open}, \xi$ and $\beta$:  
%%%%%%%%%%%%%%%%%%%%%%%%%%%%%%%%%%%%%%%%%%%%%%%%%%%%%%%%%%%%%%%%%%%%%%%%%%%
\begin{eqnarray}
  \Psi &\equiv& \int_{\rm jet}\!d\Omega'\,\psi(\Omega') \nonumber\\
       &=& -\frac{2\pi}{\beta\sin^2\xi}
           \bigg\{-2\beta\lambda(1+\cos^2\xi) \label{eq:Psi_1}\\
      &\;&\;+\left(\beta\lambda-\cos\xi\right)
           \sqrt{(\beta\cos\xi-\lambda)^2+(1-\beta^2)(1-\lambda^2)}\nonumber\\
      &\;&\;+\left(\beta\lambda+\cos\xi\right)
           \sqrt{(\beta\cos\xi+\lambda)^2+(1-\beta^2)(1-\lambda^2)}\nonumber
      \bigg\}
\end{eqnarray}
%%%%%%%%%%%%%%%%%%%%%%%%%%%%%%%%%%%%%%%%%%%%%%%%%%%%%%%%%%%%%%%%%%%%%%%%%%%
with $\lambda$ being $\cos(\theta_{\rm open}/2)$. In performing 
the integration, the range of the integral denoted by ``jet'' 
is restricted to the jet region, i.e., 
$0<\theta'<\theta_{\rm open}/2$ and $\pi-\theta_{\rm
open}/2<\theta'<\pi$.
In the ultra-relativistic limit ($\beta\to 1$), this yields
%%%%%%%%%%%%%%%%%%%%%%%%%%%%%%%%%%%%%%%%%%%%%%%%%%%%%%%%%%%%%%%%%%%%%%%%%%%
\begin{equation}
  \Psi =
  \begin{cases}
  \displaystyle 
  4\pi\lambda\left(\frac{1-|\cos\xi|}{\sin\xi}\right)^2
      &  {\rm for}~ {|\cos\xi|}> {\lambda}, \\
  \displaystyle 
  4\pi(1-\lambda)\frac{\lambda-\cos^2\xi}{\sin^2\xi} 
      &  {\rm for}~ |\cos\xi|\leq\lambda.
  \end{cases}
\label{eq:Psi_2}
\end{equation}
%%%%%%%%%%%%%%%%%%%%%%%%%%%%%%%%%%%%%%%%%%%%%%%%%%%%%%%%%%%%%%%%%%%%%%%%%%%

\begin{figure}
 \includegraphics[width=8cm]{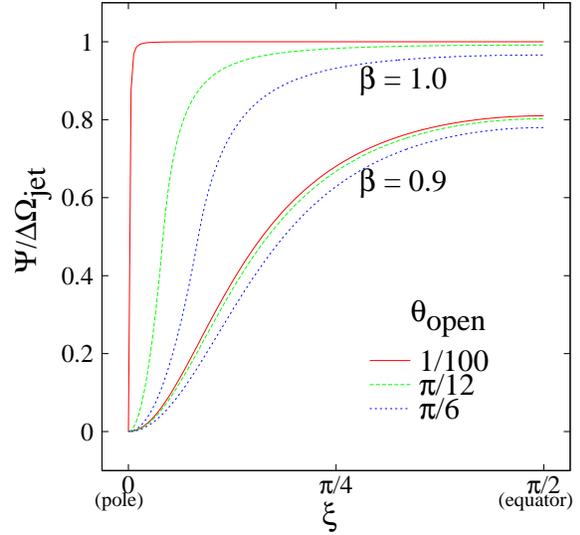}
 \caption{Viewing-angle dependence of the amplitude. The upper three 
lines represent the ultra-relativistic cases ($\beta=1.0$), and lower 
ones relativistic cases ($\beta=0.9$). We take three kinds of the 
opening-angle, $\theta_{\rm open}=1/100, \pi/12, \pi/6$.}
 \label{fig:Psi}
\end{figure}

Figure~\ref{fig:Psi} 
illustrates the viewing-angle dependence of the GW amplitude 
$\Psi/\Delta\Omega_{\rm jet}$,  
in the cases of relativistic ($\beta=0.9$) 
and ultra-relativistic limit ($\beta=1.0$). 
Figure~\ref{fig:Psi} clearly 
shows that the intensity of the GWs generated from
an axisymmetric relativistic jet
has an anti-beaming distribution. 
Compared to the relativistic case ($\beta< 1)$,  
the anti-beaming feature of ultra-relativistic jet  
is very sensitive to the opening angle, 
while the viewing-angle dependence becomes fairly weak 
for a typical value of the opening angle, $\theta_{\rm open}\leq\pi/6$ 
\citep{Frail}. 
This indicates that the GWs from the neutrino-driven GRB 
are uniformly emitted in contrast to the axisymmetric 
emission of the neutrinos itself.

Analytic expression 
(\ref{eq:h_c}) with (\ref{eq:Psi_1}) and (\ref{eq:Psi_2}) 
is one of the main result in this paper.  
It describes the gravitational waveform 
generated from a (ultra-)relativistic 
emission of neutrinos and/or matter with opening angle $\theta_{\rm open}$ 
and with the velocity $\beta$. 
With the analytic expression, one can further estimate the 
GWB spectrum in a tractable manner. Before doing this, we first apply this formula to estimate the amplitude of GW from a 
single neutrino-driven GRB in next section. 

%%%%%%%%%%%%%%%%%%%%%%%%%%%%%%%%%%%%%%%%%%%%%%%%%%%%%%%%%%%%%%%%%%%%%%%%%%%
%%%%%%%%%%%%%%%%%%%%%%%%%%%%%%%%%%%%%%%%%%%%%%%%%%%%%%%%%%%%%%%%%%%%%%%%%%%
%%%%%%%%%%%%%%%%%%%%%%%%%%%%%%%%%%%%%%%%%%%%%%%%%%%%%%%%%%%%%%%%%%%%%%%%%%%
%%%%%%%%%%%%%%%%%%%%%%%%%%%%%%%%%%%%%%%%%%%%%%%%%%%%%%%%%%%%%%%%%%%%%%%%%%%
%%%%%%%%%%%%%%%%%%%%%%%%%%%%%%%%%%%%%%%%%%%%%%%%%%%%%%%%%%%%%%%%%%%%%%%%%%%
%%%%%%%%%%%%%%%%%%%%%%%%%%%%%%%%%%%%%%%%%%%%%%%%%%%%%%%%%%%%%%%%%%%%%%%%%%%

%%%%%%%%%%%%%%%%%%%%%%%%%%%%%%%%%%%%%%%%%%%%%%%%%%%%%%%%%%%%%%%%%%%%%%%%%%
%%%%%%%%%%%%%%%%%%%%%%%%%%%%%%%%%%%%%%%%%%%%%%%%%%%%%%%%%%%%%%%%%%%%%%%%%%
\section{Gravitational waves from a Single GRB}  
\label{sec:gw-single}
%%%%%%%%%%%%%%%%%%%%%%%%%%%%%%%%%%%%%%%%%%%%%%%%%%%%%%%%%%%%%%%%%%%%%%%%%%
%%%%%%%%%%%%%%%%%%%%%%%%%%%%%%%%%%%%%%%%%%%%%%%%%%%%%%%%%%%%%%%%%%%%%%%%%%

Provided the time evolution of the luminosity $\Gamma(t)$, the analytic 
formulae (\ref{eq:h_c}) with equation (\ref{eq:Psi_1}) and/or (\ref{eq:Psi_2}) 
enable us to estimate 
the frequency dependence of the characteristic amplitude, depending 
on the parameters $(\xi, \theta_{\rm open}, E, r)$.  
In this section, for a simple but a realistic model of the luminosity 
evolution, we adopt the fitting result by \citet{Aloy} to calculate the 
GW spectrum from a single GRB. 

According to the state-of-the-art three-dimensional hydrodynamic 
simulations \citep[][]{Seti}, the fitting function is given by
%%%%%%%%%%%%%%%%%%%%%%%%%%%%%%%%%%%%%%%%%%%%%%%%%%%%%%%%%%%%%%%%%%%%%%%%%%%
\begin{equation}
 \Gamma(t) = \frac{2}{6t_{\rm f}-t_{\rm b}}
  \begin{cases}
    (t/t_{\rm b}) & 0\leq t <t_{\rm b} \\
    1             &  t_{\rm b}\leq t<t_{\rm f} \\
    (t/t_{\rm f})^{-3/2} & t_{\rm f}\leq t 
   \end{cases},
   \label{eq:lumievo}
 \end{equation}
%%%%%%%%%%%%%%%%%%%%%%%%%%%%%%%%%%%%%%%%%%%%%%%%%%%%%%%%%%%%%%%%%%%%%%%%%%%
and $\Gamma(t)=0$ for $t<0$. 
Here, the time $t_{\rm b}$ roughly corresponds to 
$t_{\rm b}\approx t_{\rm f}/3$. 
Note that the three phases of the time evolution in equation
(\ref{eq:lumievo}) 
are intimately related to the evolutionary phases of the neutrino
emission; 
i) the mass accretion, which makes the luminosity linearly growing,
ii) stationary phase after stopping
the mass fall, and iii) slow decay  
by the termination of neutrino production in the accretion
disk. These basic behaviors should be fairly insensitive to a detailed
modeling of neutrino emission. Although \citet{Aloy} only considered a
short-duration GRB in which the duration $t_\mathrm{b}$ and  
$t_\mathrm{f}$ were set to 
$0.01$ sec and $0.03$ sec, respectively, we keep to use equation
(\ref{eq:lumievo}) for a long-duration GRB.
Additionally, note that the observed duration is a few times larger  
than the duration $t_\mathrm{f}$  
due to the tail part of equation
(\ref{eq:lumievo}) at the late time. The observed duration is typically
0.2 sec for short-duration GRBs and 20 sec for long-duration GRBs
\citep{zhang}.

\begin{figure}
 \includegraphics[width=8cm]{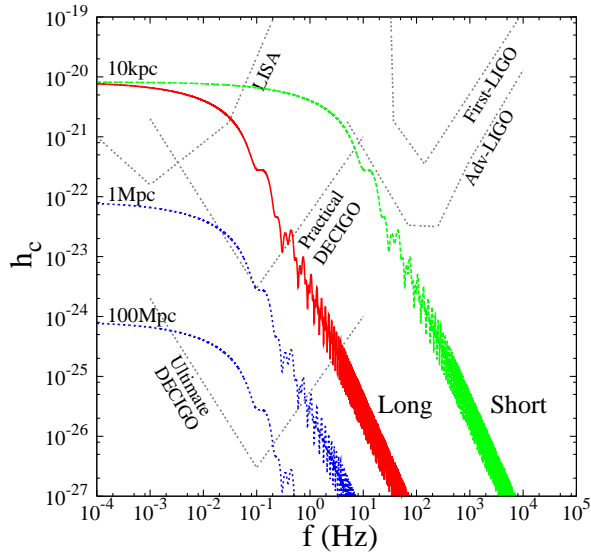}
\caption{The characteristic amplitude of GWs from a single GRB at the
 the Galactic centre (solid line). The ``Long'' duration $t_\mathrm{f}$
 is 10 sec and ``Short'' is 0.1 sec (long-dashed line). 
On the other hand, the short-dashed lines  
represent the ``Long'' duration GRB at 1Mpc (upper) 
and 100Mpc (lower). To compute the GW amplitudes, 
 we specifically set the model parameters to 
$\xi=\pi/2$, $\theta_\mathrm{open}=\pi/6, E=10^{52}$ erg.
 }
 \label{fig:sGRB}
\end{figure}

Figure~\ref{fig:sGRB} plots typical examples 
of the characteristic amplitudes for neutrino-driven GRBs. 
The solid line shows the amplitude of GWs from the 
GRB located at the Galactic centre ($r=$10kpc), 
while the short-dashed lines represent the GRBs located at $r=$1Mpc (upper)
and 100Mpc (lower). Here, we assume 
the total isotropic neutrino energy, $E=10^{52}$ erg \citep{popham},
in order to set an upper bound of the contributions 
from the neutrinos to the GW amplitudes. 
Further, extrapolating the neutrino luminosity evolution 
for the short-duration GRBs \citep{Seti} to that for the long-duration GRBs, 
GW amplitude from the long GRB was also calculated 
by simply setting $t_{\rm f}=10$ sec (solid line, 
labeled as ``Long'').

Clearly from figure~\ref{fig:sGRB}, the dominant part of 
the GWs appears at a low-frequency band and they enter 
the detectable band by future observatories. 
While the high-frequency part of GWs rapidly falls down,  
the characteristic amplitude of low-frequency GW 
asymptotically approaches constant,  
which implies 
that Fourier component of GWs $\tilde{h}(f)$ is inversely proportional 
to the frequency, i.e., $\tilde{h}\propto 1/f$. This 
behavior typically arises from a sudden change of the neutrino 
luminosity and can be deduced from the so-called zero-frequency 
limit of GW memory \citep[e.g.,][]{BT,Epstein,Buo}.

The results in figure~\ref{fig:sGRB} indicate that 
the planned space interferometers, LISA and practical DECIGO 
can easily detect the GWs from a GRB at the Galactic centre,
though the local GRB rate around the Galaxy is extremely small. 
The GRB rate would be increased as the observed volume is
increasing. For instance, we might observe a few GRB events for a decade
within 100Mpc. As shown in the figure, however, the amplitude of GWs
from such a far GRB is quite small, which may be comparable or smaller
than that of GWs generated in the inflationary epoch.
In addition, while one may detect such GW by an ultimate version of the
space interferometer whose sensitivity is limited by the quantum 
noise, the detection of GW memory needs a somewhat different 
technique from periodic and burst-like GWs, which will be an 
important subject of GW data analysis.

%%%%%%%%%%%%%%%%%%%%%%%%%%%%%%%%%%%%%%%%%%%%%%%%%%%%%%%%%%%%%%%%%%%%%%%%%%%
%%%%%%%%%%%%%%%%%%%%%%%%%%%%%%%%%%%%%%%%%%%%%%%%%%%%%%%%%%%%%%%%%%%%%%%%%%%
%%%%%%%%%%%%%%%%%%%%%%%%%%%%%%%%%%%%%%%%%%%%%%%%%%%%%%%%%%%%%%%%%%%%%%%%%%%
%%%%%%%%%%%%%%%%%%%%%%%%%%%%%%%%%%%%%%%%%%%%%%%%%%%%%%%%%%%%%%%%%%%%%%%%%%%

%%%%%%%%%%%%%%%%%%%%%%%%%%%%%%%%%%%%%%%%%%%%%%%%%%%%%%%%%%%%%%%%%%%%%%%%%%
%%%%%%%%%%%%%%%%%%%%%%%%%%%%%%%%%%%%%%%%%%%%%%%%%%%%%%%%%%%%%%%%%%%%%%%%%%
\section{Estimating the GWB from neutrino-driven GRBs}
\label{sec:gw-background}
%%%%%%%%%%%%%%%%%%%%%%%%%%%%%%%%%%%%%%%%%%%%%%%%%%%%%%%%%%%%%%%%%%%%%%%%%%
%%%%%%%%%%%%%%%%%%%%%%%%%%%%%%%%%%%%%%%%%%%%%%%%%%%%%%%%%%%%%%%%%%%%%%%%%%

We are in a position to discuss 
the contribution of GWs from the neutrino-driven GRBs to 
the background radiation. We assume that all GRBs have same
energy and opening-angle ($\theta_{\rm open}=\pi/6$), and the
viewing-angle is randomly distributed. According to 
\citet{Phinney}, the sum of the energy densities radiated by a large
number of independent GRBs at each redshift 
is given by the density parameter $\Omega_{\rm
GW}(f)\equiv \rho_c^{-1}(d\rho_{\rm GW}/d\log f)$ as:  
%%%%%%%%%%%%%%%%%%%%%%%%%%%%%%%%%%%%%%%%%%%%%%%%%%%%%%%%%%%%%%%%%%%%%%%%%%%
\begin{eqnarray}
 \Omega_{\rm GW}  &=& \frac{2\pi^2 c}{G\rho_c} \int_0^\infty \!dz\,
                    \frac{R_{\rm GRB}(z)}{1+z}
                    \frac{1}{(1+z)H(z)}
\nonumber \\
            && \times r^2 f^3
                    \langle|\widetilde{h}_+(f)|^2\rangle_{\xi,t_{\rm f}}
              \Big|_{t_{\rm f}\to t_{\rm f}/(1+z)\atop f\to f(1+z)},
\label{eq:OmegaGW}
\end{eqnarray}
%%%%%%%%%%%%%%%%%%%%%%%%%%%%%%%%%%%%%%%%%%%%%%%%%%%%%%%%%%%%%%%%%%%%%%%%%%%
with $R_{\rm GRB}(z)$ being redshift distribution of GRBs. $H(z)$
is the Hubble parameter.
Here the quantity $\langle|\widetilde{h}_+(f)|^2\rangle_{\xi,t_{\rm f}}$
represents the averaged value of the amplitude 
$|\widetilde{h}_+(f)|^2$ over the
viewing-angle $\xi$ and the duration $t_{\rm f}$. 
Note that the combination 
$r^2f^2\langle|\widetilde{h}_+(f)|^2\rangle_{\xi,t_{\rm f}}$  
becomes independent of the distance $r$, and the explicit 
dependence on the redshift $z$ is eliminated
after replacing $(f,t_{\rm f})$ with 
$(f(1+z),t_{\rm f}/(1+z))$.

In order to quantify the amplitude of GWB, we need the redshift
distribution of GRBs.  For a crude 
estimate of the amplitude of $\Omega_{\rm GW}$, 
we use the redshift evolution inferred from the 
$E_p$-luminosity correlation \citep{Amati,Yone}: 
%%%%%%%%%%%%%%%%%%%%%%%%%%%%%%%%%%%%%%%%%%%%%%%%%%%%%%%%%%%%%%%%%%%%%%%%%%%
\begin{equation}
 R_{\rm GRB}(z) \propto
  \begin{cases}
   (1+z)^{6.0\pm1.4} & z < 1 \\
   (1+z)^{0.4\pm0.2} & z > 1
  \end{cases}. \label{eq:redshift}
\end{equation}
%%%%%%%%%%%%%%%%%%%%%%%%%%%%%%%%%%%%%%%%%%%%%%%%%%%%%%%%%%%%%%%%%%%%%%%%%%%
As for the normalization factor $R_{\rm GRB}(0)$, 
the local observation by BATSE indicates 
$10^{-7}$ per year per a galaxy, but, the jets are highly aligned and
the true GRB rate must be larger.
Then, the local GRB rate becomes $\sim10^{-5}$
per year per a galaxy \citep{Frail}, or equivalently, 
%%%%%%%%%%%%%%%%%%%%%%%%%%%%%%%%%%%%%%%%%%%%%%%%%%%%%%%%%%%%%%%%%%%%%%%%%%%
\begin{equation}
 R_{\rm GRB}(0) =  25 \sim  250\;{\rm Gpc}^{-3} {\rm yr}^{-1}. 
 \label{eq:local}
\end{equation}
%%%%%%%%%%%%%%%%%%%%%%%%%%%%%%%%%%%%%%%%%%%%%%%%%%%%%%%%%%%%%%%%%%%%%%%%%%%

Assuming that the orientation of jets is random, 
the ensemble average 
$\langle|\widetilde{h}_+(f)|^2\rangle_{\xi,t_{\rm f}}$
in equation (\ref{eq:OmegaGW}) is evaluated separately, namely, 
$\langle|\widetilde{h}_+(f)|^2\rangle_{\xi,t_{\rm f}}\propto
\langle|\Psi|^2\rangle_\xi\langle|\widetilde{\Gamma}(t)|^2
\rangle_{t_{\rm f}}$. For the average over the viewing-angle $\xi$, one finds 
%%%%%%%%%%%%%%%%%%%%%%%%%%%%%%%%%%%%%%%%%%%%%%%%%%%%%%%%%%%%%%%%%%%%%%%%%%%
\begin{equation}
 \langle|\Psi|^2\rangle_\xi 
    = \frac{1}{4\pi}\int_0^{\pi}\!\Psi(\xi)^2\;2\pi\sin\xi\,d\xi.
\end{equation}
%%%%%%%%%%%%%%%%%%%%%%%%%%%%%%%%%%%%%%%%%%%%%%%%%%%%%%%%%%%%%%%%%%%%%%%%%%%
The integration can be done analytically, but the result is rather 
complicated. For a typical set of parameters 
$(\theta_{\rm open}, \beta)=(\pi/6,1)$, this gives 0.154.  
On the other hand, for the average of the GW amplitude over the 
duration, the distribution function for the duration
of GRBs, $T(t_{\rm f})$ must be taken into account. 
Here, we use the data set of durations ($T_{90}$) taken from the BATSE
4B catalog\footnote{
http://www.batse.msfc.nasa.gov/batse/grb/catalog/4b/4br$\_$duration.html.}
adopting a logarithmic binning
with $\Delta\log_{10}t_{\rm f}=0.1$.
The resultant distribution becomes a bimodal distribution over the range 
0.015 sec $\leq t_{\rm f}\leq $ 670 sec. Since the function 
$\widetilde{\Gamma}(f)$ is the only quantity that explicitly depends on 
the duration $t_{\rm f}$, the ensemble average of the GW amplitude 
may be independently evaluated as:  
%%%%%%%%%%%%%%%%%%%%%%%%%%%%%%%%%%%%%%%%%%%%%%%%%%%%%%%%%%%%%%%%%%%%%%%%%%%
\begin{equation}
 \langle|\widetilde{\Gamma}(t)|^2\rangle_{t_{\rm f}} =
  \frac{\int\!|\widetilde{\Gamma}(t)|^2T(t_{\rm f})\,dt_{\rm
  f}}{\int\!T(t_{\rm f})\,dt_{\rm f}}.
\label{eq:average_tf}
\end{equation}
%%%%%%%%%%%%%%%%%%%%%%%%%%%%%%%%%%%%%%%%%%%%%%%%%%%%%%%%%%%%%%%%%%%%%%%%%%%

%%%%%%%%%%%%%%%%%%%%%%%%%%%%%%%%%%%%%%%%%%%%%%%%%%%%%%%%%%%%%%%%%%%%%%%%%%%
%%%%%%%%%%%%%%%%%%%%%%%%%%%%%%%%%%%%%%%%%%%%%%%%%%%%%%%%%%%%%%%%%%%%%%%%%%%
\begin{figure}
 \includegraphics[width=8cm]{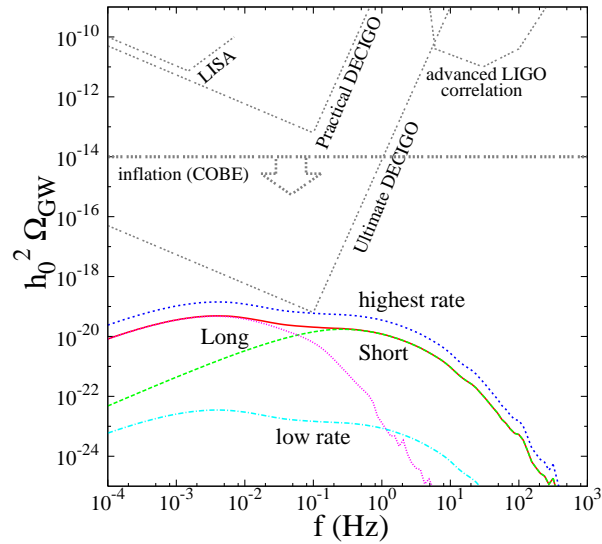}
 \caption{The energy density parameter of GWB. The solid line represents
 the contribution from neutrinos of all GRBs. The dotted line
 represents one from ``Long''-duration GRBs ($t_\mathrm{f}>2$ sec), and
 the long-dashed line one from ``Short''-duration GRBs
 ($t_\mathrm{f}<2$ sec). The dotted-dashed line represents
 the case of $R_\mathrm{GRB}(0)=25\;{\rm Gpc}^{-3} {\rm
 yr}^{-1}$,$R_\mathrm{GRB}\propto (1+z)^{4.6}, (1+z)^{0.2}$  (``low 
 rate''). 
 On the other hand, the short-dashed line represents the case of
 highest GRB rates, $R_\mathrm{GRB}(0)=250\;{\rm Gpc}^{-3} {\rm yr}^{-1}$,
 $R_\mathrm{GRB}\propto (1+z)^{7.4}, (1+z)^{0.6}$ in the cases of $z<1$
 and $z>1$, respectively. The 
 horizontal dotted line shows the GWB from the
 inflationary epoch constrained by COBE observations.}
 \label{fig:GWB}
\end{figure}
%%%%%%%%%%%%%%%%%%%%%%%%%%%%%%%%%%%%%%%%%%%%%%%%%%%%%%%%%%%%%%%%%%%%%%%%%%%
%%%%%%%%%%%%%%%%%%%%%%%%%%%%%%%%%%%%%%%%%%%%%%%%%%%%%%%%%%%%%%%%%%%%%%%%%%%

Figure~\ref{fig:GWB} shows the density parameter of GWB generated from
neutrino-driven GRBs (solid line), together with the sensitivity curves
for future missions.  
To plot the result, we adopt the flat cosmological
model with the density parameters $\Omega_M = 0.3$ and $\Omega_\Lambda =
0.7$. The Hubble parameter is given by
%%%%%%%%%%%%%%%%%%%%%%%%%%%%%%%%%%%%%%%%%%%%%%%%%%%%%%%%%%%%%%%%%%%%%%%%%%%
\begin{equation}
H(z) = H_0\sqrt{\Omega_M(1+z)^3+\Omega_\Lambda}
\end{equation} 
%%%%%%%%%%%%%%%%%%%%%%%%%%%%%%%%%%%%%%%%%%%%%%%%%%%%%%%%%%%%%%%%%%%%%%%%%%%
with the present Hubble parameter $H_0$ being $72$ km s$^{-1}$/Mpc. 
Note that when evaluating the expression (\ref{eq:OmegaGW}), 
the range of the integral is restricted to $0\leq z\leq 20$, 
since most of the contribution ($\sim 80\%$) to the integral comes from
the low redshift GRBs $z\la 4$.

Figure ~\ref{fig:GWB} shows that the GWB 
is broadly distributed over the low-frequency band, 
$10^{-3}{\rm Hz}<f<10^1{\rm Hz}$ with amplitude
$h_0^2\Omega_{\rm GW}\sim10^{-20}$  
and has almost the flat spectrum with small bimodal bump.  
To see the contribution 
from short-duration GRBs and long-duration GRBs separately, the 
ensemble average over the duration (\ref{eq:average_tf}) is 
artificially divided into the short duration ($t_f\leq2$ sec) and 
the long duration ($t_f>2$ sec), the results of which are 
respectively shown in dashed and dotted lines. 
One then immediately deduces that bimodal shape of the 
GWB spectrum has originated from a distribution of duration.  
This is mainly because the low-frequency part of the 
gravitational waveform shown in figure \ref{fig:sGRB} 
becomes featureless due to the zero-frequency limit of the 
memory effect.  Although the present result was derived from a 
specific choice of luminosity evolution (\ref{eq:lumievo}), 
the characteristic behavior seen in figure \ref{fig:GWB} 
would remain unchanged as long as the collapsar model 
is applicable to both short- and long-duration GRBs.

Figure \ref{fig:GWB} indicates that the GWB from neutrino-driven GRBs is
much smaller than the upper limit of GWB generated in the inflationary
epoch constrained by the COBE observations (the horizontal thick

dotted-line) and also below the detection limit of future space
missions.  
Note that this tiny amplitude can be also deduced from 
an extrapolation of the discussion by \citet{Buo} based on the following 
points; i) the GRB rate is roughly four orders of magnitude smaller 
than the rate 
of supernovae and ii) 
the total released energy by neutrinos used in this paper 
is thirty times smaller than theirs. 
According to \citet{Buo}, the maximum amplitude of GWB from cosmological supernovae is estimated as 
$\Omega_{\rm GW}^{\rm SN} \approx 10^{-12}$ (see Fig. 4 of their 
paper). Since the density 
parameter $\Omega_{\rm GW}$ is proportional to 
$R_\text{GRB}(0)  E_\nu^2$ (see Eqs.(\ref{eq:h_c}) and (\ref{eq:OmegaGW})), 
we roughly obtain $\Omega_{\rm GW}^{\rm GRB} \approx 10^{-19}$ 
in the cosmological GRB case. In this sense, small amplitude of GWB may 
be a natural outcome from the rare even rate of GRB.    
On the other hand, nearly flat spectrum of GWB in figure \ref{fig:GWB} 
cannot be simply explained by an argument based on the 
the supernovae case, which may be a unique characteristic of the 
GRB case.

As examined in figure \ref{fig:GWB} (thick-dotted and 
dot-dashed lines), the uncertainty of the normalization in the GRB rate
(\ref{eq:local}) does not change the conclusion. 
In other words, the GWB generated from a cosmological population of GRBs
does not become a preventer of the primordial GWB generated in the
inflationary epoch.

%%%%%%%%%%%%%%%%%%%%%%%%%%%%%%%%%%%%%%%%%%%%%%%%%%%%%%%%%%%%%%%%%%%%%%%%%%
%%%%%%%%%%%%%%%%%%%%%%%%%%%%%%%%%%%%%%%%%%%%%%%%%%%%%%%%%%%%%%%%%%%%%%%%%%
\section{Summary and Discussion} 
\label{sec:summary}
%%%%%%%%%%%%%%%%%%%%%%%%%%%%%%%%%%%%%%%%%%%%%%%%%%%%%%%%%%%%%%%%%%%%%%%%%%
%%%%%%%%%%%%%%%%%%%%%%%%%%%%%%%%%%%%%%%%%%%%%%%%%%%%%%%%%%%%%%%%%%%%%%%%%%

We have discussed the GWs from 
axisymmetric relativistic jets of neutrino-driven GRBs.
In order to
estimate the amplitude of the GWs, we derived the analytic formulae
for the gravitational waveform from 
(ultra-)relativistic energy flow. The analytic expression given in 
equation (\ref{eq:h_c}) with (\ref{eq:Psi_1}) and (\ref{eq:Psi_2}) 
depends on the viewing-angle $\xi$, 
the opening-angle of the jet $\theta_{\rm open}$, 
and the velocity of the flow $\beta$, as well as the time evolution of 
luminosity $\Gamma(t)$ (or $\tilde{\Gamma}(f)$). 
Following the fitting result
obtained from the state-of-the-art numerical
simulations of collapsars, the amplitudes of the GWs from a single GRB are 
estimated and are found that 
within the detection limits for the space-based laser interferometers
like LISA and DECIGO/BBO, if a GRB occurs at the region within 1Mpc. 
Since the released energy in the jet region by neutrinos can be
greater than that by matter, the low-frequency GWs  
generated by neutrinos seem to 
dominate the one by matter studied by \citet{Sago}. 
Note that, even if the amplitudes of 
the GW memory overcome the 
sensitivity curves of interferometers in the spectra, this does not 
directly imply the detection of the signals. 
The detectability should be discussed 
with a specific data analysis technique for the memory effect. 
On the other hand, the GWB from a cosmological 
population of GRBs are obtained by summing up the individual GW of the GRBs, 
which turns out to be sufficiently smaller than those from the
inflationary epoch and a cosmological population of
supernovae \citep[][]{Buo}.

While the sufficiently small amplitude of the GWB would be true, there remain
several uncertainties in predicting a precise waveform of the GW from a
single GRB jet. In particular, when estimating the GW from a long 
GRB, we have extrapolated to use the luminosity 
evolution of neutrinos for a short GRB. 
Since the efficient mechanism to trigger the long burst is still
under debate, one should continue to check the present GRB model in more
quantitative way. As a prelude to
more realistic luminosity evolutions in the collapsar models,  which requires
general relativistic multidimensional radiation hydrodynamic
simulations, our findings obtained in a semi-analytic manner should be
the very first step towards the predictions of the GWs from the
collapsars and the following formation of GRBs. 

\section*{Acknowledgements}

We thank S. Yamada and K. Sato for informative
discussions. This work was supported in part by the Japan Society for
Promotion of Science(JSPS) Research Fellowships (TH, KK, HK) and a
Grant-in-Aid for Scientific Research from the JSPS(AT, No.14740157). 

%\clearpage
%%%%%%%%%%%%%%%%%%%%%%%%%%%%%%%%%%%%%%%%%%%%%%%%%%%%%%%%%%%%%%%%%%%%%%%%%%
%%%%%%%%%%%%%%%%%%%%%%%%%%%%%%%%%%%%%%%%%%%%%%%%%%%%%%%%%%%%%%%%%%%%%%%%%%
%%%%%%%%%%%%%%%%%%%%%%%%%%%%%%%%%%%%%%%%%%%%%%%%%%%%%%%%%%%%%%%%%%%%%%%%%%

%%%%%%%%%%%%%%%%%%%%%%%%%%%%%%%%%%%%%%%%%%%%%%%%%%%%%%%%%%%%%%%%%%%%%%%%%%
%%%%%%%%%%%%%%%%%%%%%%%%%%%%%%%%%%%%%%%%%%%%%%%%%%%%%%%%%%%%%%%%%%%%%%%%%%
%%%%%%%%%%%%%%%%%%%%%%%%%%%%%%%%%%%%%%%%%%%%%%%%%%%%%%%%%%%%%%%%%%%%%%%%%%

\label{lastpage}
\end{document}